# DBSCAN for nonlinear equalization in high-capacity multi-carrier optical communications

**Elias Giacoumidis** [1,*,†], **Yi Lin** [1,†], **and Liam P. Barry**[1]

[1]  Dublin City University, School of Electronic Engineering, Radio and Optical Communications Lab, Glasnevin 9, Dublin, Ireland.

*  Correspondence: elias.giacoumidis@dcu.ie

†  Equally contributed to this work.

**Abstract:** Coherent optical multi-carrier communications have recently dominated metro-regional and long-haul optical communications for signal capacities per wavelength higher than 20 Gbit/sec. However, the major obstacle of fiber-optic networks involving high-bandwidth coherent multi-carrier signals such as coherent optical orthogonal frequency-division multiplexing (CO-OFDM) is the fiber-induced nonlinearity and the parametric noise amplification from cascaded optical amplifiers which results in significant nonlinear distortion among subcarriers. Here, we present the first nonlinear equalizer in optical communications using the traditional Density-Based Spatial Clustering of Applications with Noise (DBSCAN) algorithm and a novel modified version of DBSCAN which combines K-means clustering on the noisy "un-clustered" symbols. For a 24.72 Gbit/sec differential quaternary phase-shift keying (DQPSK) CO-OFDM system, the modified DBSCAN can increase the signal quality-factor by up to 2.158 dB compared to linear equalization at 500 km of transmission. The modified DBSCAN slightly outperforms the traditional DBSCAN, fuzzy-logic C-means, hierarchical and conventional K-means clustering at high launched optical powers.

## 1. Introduction

A few unsupervised machine learning algorithms have been recently implemented in optical fiber communications for training-data-free nonlinear equalization. These unsupervised algorithms included for example fuzzy logic C-means [1], K-means/K-nearest-neighbors [1, 2], hierarchical [1], affinity propagation [3] and Gaussian mixture [4] clustering. In modern spectral-efficient multi-carrier modulation schemes such coherent optical orthogonal frequency-division multiplexing (CO-OFDM) [5, 6], the aforementioned algorithms have shown great potential in tackling both deterministic and stochastic nonlinearities for long-haul transmission, such as the parametric noise amplification from cascaded optical amplifiers. In particular, affinity propagation clustering based nonlinear equalization outperformed in signal quality (Q)-factor benchmark clustering algorithms [3]. However, the computational complexity of affinity propagation was estimated to be very high [7] and therefore alternative clustering algorithms are required that can be potentially considered for real-time optical communications.

Density-Based Spatial Clustering of Applications with Noise (DBSCAN) is an unsupervised machine learning algorithm that was first proposed by Ester et al. [8] for clustering analysis based on the density approach. DBSCAN has been adopted for various applications such as in data mining [8] and image processing [9] due to the attractive ability of finding clusters of any shape, identifying noisy points effectively, and supporting spatial databases. DBSCAN has been implemented in optical domain only for modulation format recognition [10] and for nonlinear equalization in visible light communications systems [11]. In this work, we implement the first DBSCAN-based nonlinear equalizer for high-capacity optical fiber communications systems. More specific, the traditional DBSCAN (referred as conventional DBSCAN throughout this work) is implemented for CO-OFDM; while also a novel modified DBSCAN algorithm is presented, in which the "un-clustered" noisy points are further processed with K-means. We numerically demonstrate that the modified DBSCAN can increase the Q-factor compared to linear equalization by up to 2.158 dB for a 24.72 Gbit/sec differential quaternary phase-shift keying (DQPSK) CO-OFDM system at 500 km of standard single-mode fibre transmission (SSMF). We also show that modified DBSCAN slightly outperforms the traditional DBSCAN, conventional K-means, fuzzy-logic C-means and hierarchical clustering at high launched optical powers.

## 2. Conventional and modified DBSCAN

Density based clustering algorithms make an assumption that clusters are densed regions in space separated by regions of lower density [8]. A densed cluster is a region which is "density connected", i.e. the density of points in that region is greater than a minimum [9]. Since these algorithms expand clusters based on dense connectivity, they can find clusters of arbitrary shapes [9, 10]. DBSCAN is an example of density-based clustering algorithms that deals with stochastic-noisy data. The main task of DBSCAN is to search for dense areas and to expand these recursively to find arbitrarily densed-shaped clusters. The two main parameters of DBSCAN are the $\varepsilon$ ('Epsilon') and the 'minimum points'. The $\varepsilon$ defines the radius of the "neighborhood region" while the 'minimum points' defines the minimum number of points that should be contained within that neighborhood. Every neighbor point, which its $\varepsilon$-neighborhood, contains a predefined number of points, while the cluster is also expanded to contain its neighbors [12]. However, for the unallocated points, if the number of points in the neighborhood is less than a predefined threshold, the point is designated to be noisy [12]. The noisy data in conventional DBSCAN are not further processed by machine learning and for this reason we proposed in this work to apply K-means as a 2nd stage clustering only for these noisy symbols.

A typical example for conventional DBSCAN is shown below in Fig. 1 for the case when the number of minimum points is equal to 4. In Fig. 1 we assume the following 6 assumptions [12]:

1. **Epsilon neighborhood (N$\varepsilon$)**: A set of all symbols within a distance '$\varepsilon$'.
2. **Core point**: A point that has at least a 'minimum point' (including itself) within its N$\varepsilon$.
3. **Direct Density Reachable**: A point $q$ is directly density reachable from a point $p$, if $p$ is core point and $q \in$ N$\varepsilon$.
4. **Density Reachable**: Two points are density reachable if there is a chain of 'direct density reachable' points that link these two points.
5. **Border Point**: Point that is 'direct density reachable' but not a core point.
6. **Noise**: Points not belonging to any point's N$\varepsilon$.

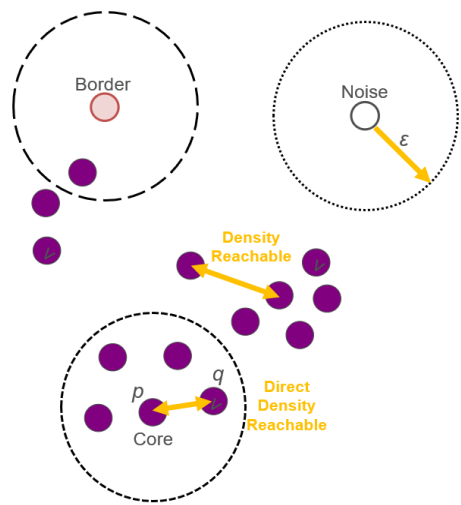

**Figure 1.** Example of DBSCAN when the number of minimum points is equal to 4.

The steps related to the conventional and modified DBSCAN are listed below, where the algorithm converges until all symbols of the DQPSK-CO-OFDM have been allocated to a cluster or labelled as 'noise' only if conventional DBSCAN is considered [8-12]:

1. Randomly select a symbol *p* (referred to Fig. 1) in the constellation map.

2. Retrieve all symbols directly density-reachable from *p* that satisfy the condition of the radius ε limits.

3. If the symbol *p* is a core point, a cluster is formed. Search recursively and find all its density connected points and assign them to the same cluster as *p*.

4. If *p* is not a core point, the DBSCAN algorithm "scans" for the rest unvisited symbols.

5. Symbols that are un-clustered are labelled as zero symbols ("noisy symbols") where linear equalization is performed only on these symbols; and then the conventional DBSCAN algorithm stops.

6. This <u>additional step</u> is only taken into account for the modified DBSCAN where K-means clustering is activated for the "noisy symbols" using the Lloyd's algorithm [1]:

   a. *Assignment*: Allocate each observation to the cluster whose mean has the least squared Euclidean distance ("nearest" mean) [1, 13].

   b. *Update*: Calculate the new means to be the centroids of the observations in the new clusters [13]. K-means algorithm converges when the assignments do not change.

## 3. Simulated CO-OFDM system equipped with DBSCAN

The CO-OFDM system incorporating the SSMF and the machine learning-based nonlinear equalizers was numerical demonstrated in Matlab® with procedures similar to these reported in Refs. [1, 14]. The adopted single-polarization/single-channel CO-OFDM system considered ideal IQ modulation and homodyne coherent reception. 128 subcarriers were employed with 400 symbols-per-subcarrier, while the cyclic prefix (CP) length [15] was taken at 10% to eliminate inter-symbol interference from linear effects such as polarization-mode dispersion (PMD). Phase noise from the frequency offset between the transmitter laser and receiver local oscillator was not considered in order to isolate stochastic and deterministic nonlinear phenomena, and hence frequency offset compensation was not required. However, digital-to-analogue/analogue-to-digital converter (DAC/ADC) clipping ratio and quantization were taken into account and set to 13 dB and 10-bits, respectively, which has negligible impact on OFDM performance for a subcarrier number > 32 [16, 17]. The sampling rate was set at 12.5 GSamples/sec and the raw signal-bit rate at 24.72 Gbit/sec. To avoid cycle slips, the detection was differential using QPSK (i.e. DQPSK). A typical block-diagram of the simulated coherent receiver including the CO-OFDM digital signal processing (DSP) unit is shown in Fig. 2. After the optical-to-electrical conversion the signal in Fig. 2 is parallelized, the CP is removed and then the fast Fourier transform is placed to demultiplex the electronic subcarriers. Afterwards, linear equalization is employed which involves pilot-assisted channel estimation and chromatic dispersion (CD) compensation. In the final stage, just before decoding and serialization, the DBSCAN algorithm (conventional and modified) is inserted to perform nonlinear equalization. It should be noted that on the same block, other clustering algorithms were also tested for comparison namely the fuzzy-logic C-means, K-means, and hierarchical clustering with procedures identical to Ref. [1]. In the transmission-link, an Erbium-doped fiber amplifier (EDFA) was inserted after a 100 km span for a total 500 km transmission. The EDFA noise was modelled as additive white-Gaussian noise and the noise figure was 5.5 dB. The optical fiber for single-polarization transmission was modelled using the pseudo-spectral split-step Fourier method which solves the nonlinear Schrödinger equation. The adopted SSMF parameters for this work are the following: fiber nonlinear Kerr parameter, CD, CD-slope, fiber loss, and PMD coefficient of 1.1 $W^{-1}$ $km^{-1}$, 16 ps $nm^{-1}$ $km^{-1}$, 0.06 ps $km^{-1}$ $(nm^2)^{-1}$, 0.2 dB $km^{-1}$ and 0.1 ps $(km^{0.5})^{-1}$, respectively. Finally, in this work the total OFDM subcarriers' bit-error-rate (BER) by error counting and Q-factor (=$20\log_{10}[\sqrt{2}erfc^{-1}(2BER)]$) are the crucial parameters under investigation.

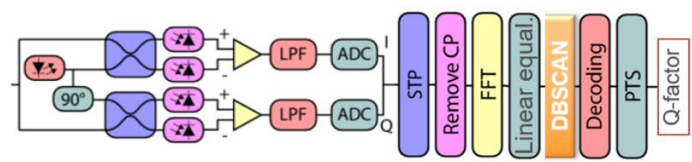

**LPF:** low-pass filter
**ADC:** analogue-to-digital converter
**STP/PTS:** serial-to-parallel/parallel-to-serial
**CP:** cyclic prefix
**FFT:** fast-Fourier transform (FFT)

**Figure 2.** Block-diagram of adopted coherent optical OFDM receiver incorporating DBSCAN for nonlinear equalization.

## 4. DBSCAN optimization and results

In order to optimize DBSCAN, the $\varepsilon$ and minimum points can be adjusted to find the lowest BER (and therefore the highest Q-factor). An example for our DQPSK-CO-OFDM system at 4 dBm of launched optical power (LOP) is shown below in Figs. 3,4, where these parameters are tuned in terms of the BER and the output number of clusters from the modified DBSCAN.

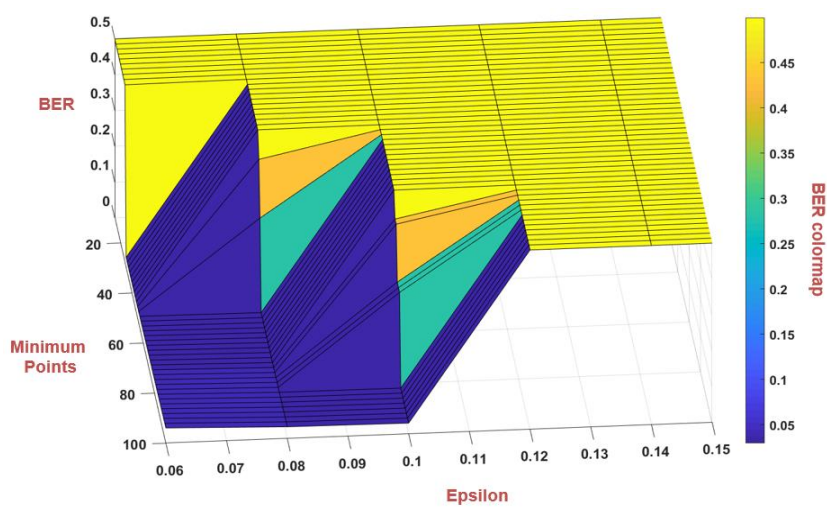

**Figure 3.** Evolution of minimum points and Epsilon ($\varepsilon$) for modified DBSCAN in terms of bit-error-rate (BER) when considering a DQPSK-CO-OFDM system at 500 km transmission and 4 dBm of launched optical power (LOP).

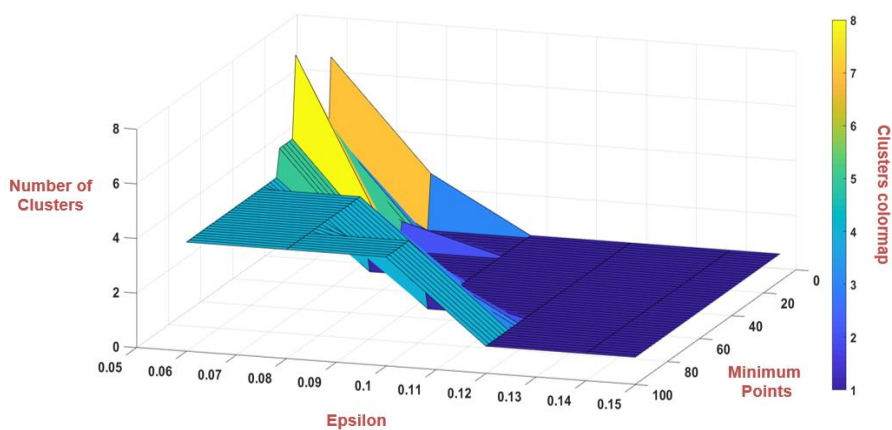

**Figure 4.** Evolution of minimum points and $\varepsilon$ for modified DBSCAN in terms of output number of clusters when considering a DQPSK-CO-OFDM system at 500 km transmission and 4 dBm of LOP.

In Fig. 3, the optimum BER is found around $0.08 < \varepsilon < 0.1$ and when the minimum points are > 80; while in Fig. 4 it is shown that the modified DBSCAN gives a wrong number of DQPSK output clusters (i.e. larger or smaller than 4) outside the aforementioned range.

Fig. 5 depicts our simulation results in terms of Q-factor vs. LOP for the adopted CO-OFDM system at 500 km of SSMF transmission. Comparisons are also made between conventional and modified DBSCAN as well as linear equalization, K-means, fuzzy-logic C-means, and hierarchical clustering. Fig. 5 reveals that DBSCAN can increase the Q-factor compared to linear equalization by up to 2.158 dB. The modified DBSCAN slightly outperforms traditional DBSCAN, fuzzy-logic C-means, hierarchical and conventional K-means clustering at high launched optical powers. In particular, at a LOP of 6 dBm the modified DBSCAN outperforms its conventional version by 0.36 dB in Q-factor. Results reveal that DBSCAN is a good candidate for fiber nonlinearity and parametric noise amplification compensation, considering that it outperforms for the whole range of LOPs. However, we believe that at low LOPs, DBSCAN (and the other machine learning clustering algorithms) performance benefit is also attributed to the fact it can make better linear symbol decisions thus providing a very effective soft-decision decoder. Finally, we should note that within the range -15 to 0 dBm of LOPs we had an error-free performance due to the limited number of symbols used to reduce simulation time. However, it is envisaged that for these power levels a similar performance benefit will be provided by DBSCAN.

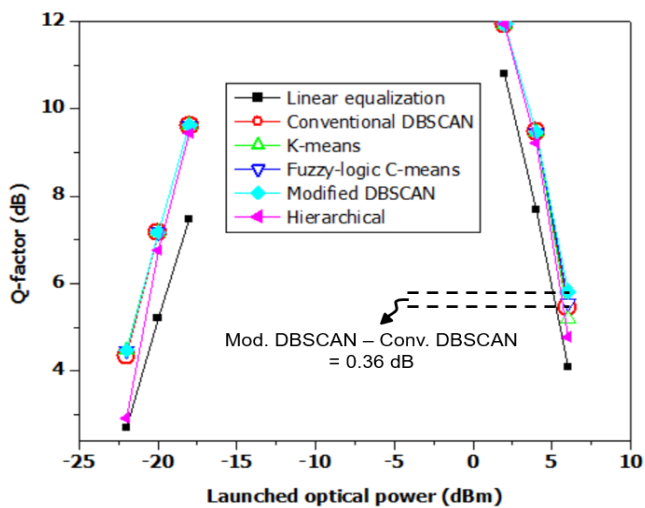

**Figure 5.** Performance comparison between different clustering algorithms in terms of Q-factor for a wide range of LOPs, considering a DQPSK-CO-OFDM system at 500 km of fiber transmission.

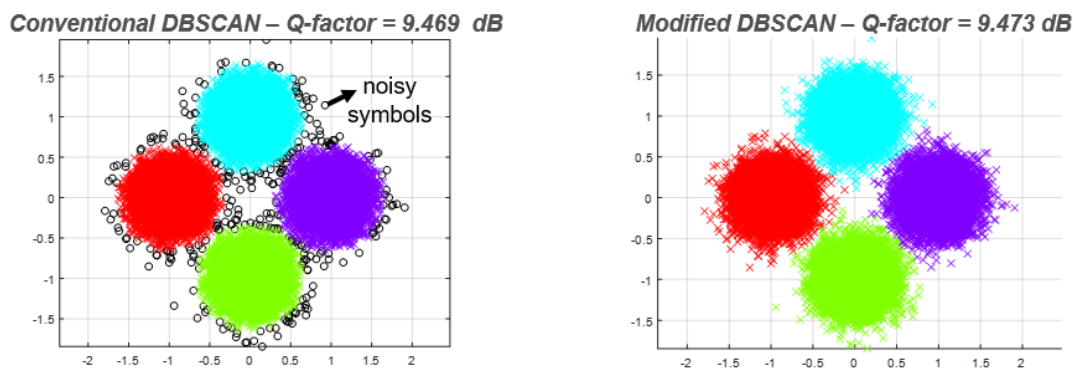

**Figure 6.** Received DQPSK constellation diagrams for conventional and modified DBSCAN at a LOP of 4 dBm.

In Figs. 6-8, we provide examples of received DQPSK-CO-OFDM constellation diagrams for all machine learning clustering algorithms under test at 4 and -20 dBm of LOP, where it is clearly presented the way these algorithms cluster the OFDM symbols. For instance, the modified DBSCAN, fuzzy-logic C-means and hierarchical clustering execute overlapping (soft) clustering while K-means and the conventional DBSCAN provide exclusive (hard) clustering.

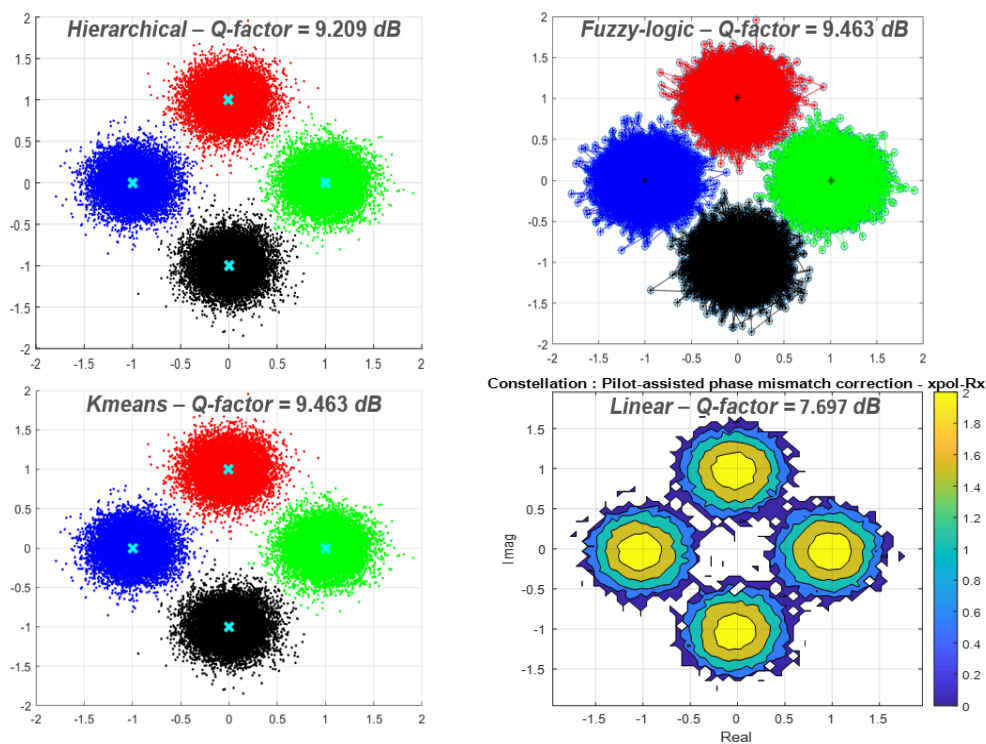

**Figure 7.** Received DQPSK constellation diagrams for hierarchical, fuzzy-logic C-means, K-means and linear equalization at a LOP of 4 dBm.

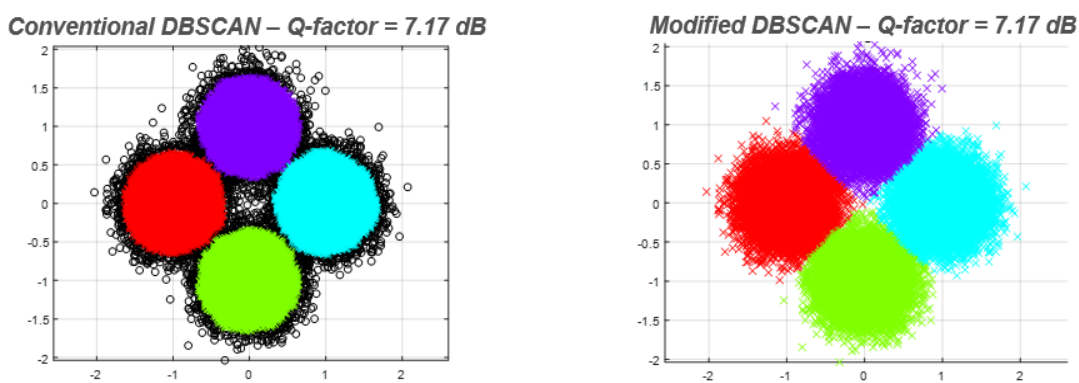

**Figure 8.** Received DQPSK constellation diagrams for conventional and modified DBSCAN at a LOP of -20 dBm.

## 5. Discussion

We numerically demonstrated the first DBSCAN-based nonlinear equalizer in optical communications using a DQPSK-CO-OFDM system for 500 km of transmission. We highlighted the performance benefits of a novel modified DBSCAN over its conventional form and compared with benchmark machine learning clustering approaches such as K-means, fuzzy-logic C-means and hierarchical clustering. We assume the marginal performance benefit provided by DBSCAN is due to the Gaussian-like circular noisy clusters; and we believe the developed algorithm should be more compatible for noisy clusters of elliptical form such as in that observed in wavelength conversion based coherent optical systems [18, 19]. We finally trust DBSCAN could be very useful for very-high spectral efficient modulation techniques such as Fast-OFDM [20-23] and in single-carrier modulated optical systems [24].

# 6. References

1. E. Giacoumidis, A. Matin, J. L. Wei, N. J. Doran, L. P. Barry, and X. Wang, "Blind Nonlinearity Equalization by Machine Learning based Clustering for Single- and Multi-Channel Coherent Optical OFDM", J. of Lightw. Techn., vol. 36, no. 3, pp. 721–727, 2018.

2. D. Wang et al., "Nonlinearity mitigation using a machine learning detector based on k-nearest neighbors", Photon. Technol. Lett., vol. 28, no.19, pp. 2102–2105, 2016.

3. E. Giacoumidis, I. Aldaya, J. L. Wei, C. Sánchez-Costa, H. Mrabet, and L. P. Barry, "Affinity propagation clustering for blind nonlinearity compensation in coherent optical OFDM", In Proc. CLEO, 2018, San Jose, CA, USA, p. STh1C.5.

4. D. Zibar et al., "Nonlinear impairment compensation using expectation maximization for dispersion managed and unmanaged PDM 16-QAM transmission", Opt. Exp., vol. 20, no. 26, pp. B181–B196, 2012.

5. W. Shieh and I. Djordjevic (2010). OFDM for Optical Communications, (Elsevier) ISBN: 9780080952062.

6. J. Armstrong, "OFDM for optical communications", J. of Lightw. Techn., vol. 27, no. 3, 189–204, 2009.

7. E. Giacoumidis, Y. Lin, J. L. Wei, I. Aldaya, A. Tsokanos, and L. P. Barry, "Harnessing machine learning for fiber-induced nonlinearity mitigation in long-haul coherent optical OFDM", Future Internet, vol. 11, no. 2, 2019.

8. M. Ester, H. P. Kriegel, J. Sander, and X. Xu, "A Density-Based Algorithm for Discovering Clusters in Large Spatial Databases with Noise," in Second International Conference on Knowledge Discovery and Data Mining (1996), pp. 226–231.

9. J. Shen, X. Hao, Z. Liang, Yu Liu, W. Wang, and L. Shao, "Real-Time Superpixel Segmentation by DBSCAN Clustering Algorithm", IEEE Trans. on Image Proc., vol. 25, no. 12, pp. 5933–5942, 2016.

10. R. Boada, R. Borkowski, and I. T. Monroy, "Clustering algorithms for Stokes space modulation format recognition", Opt. Exp., vol. 23, no. 12, pp. 15521–15531, 2015.

11. X. Lu, L. Qiao, Y. Zhou, W. Yu, and N. Chi, "An I-Q-Time 3-dimensional post-equalization algorithm based on DBSCAN of machine learning in CAP VLC system", Opt. Comms., vol. 430, pp. 299–303, 2019.

12. V. Bharti, Computer Science Dept., GC, CUNY. Lecture notes: http://www.cs.csi.cuny.edu/~gu/teaching/courses/csc76010/slides/Clustering%20Algorithm%20by%20Vishal.pdf

13. D. MacKay (2003). "Chapter 20. An Example Inference Task: Clustering" (PDF). Information Theory, Inference and Learning Algorithms. Cambridge University Press. pp. 28, 292. ISBN 0-521-64298-1. MR 2012999.

14. E. Giacoumidis, S. T. Le, I. D. Phillips, A. D. Ellis, and N. J. Doran, "Dual-polarization Multi-band OFDM Signals for Next Generation Core Networks", In Proc. CSNDSP, 23-25 July 2014, Manchester, UK.

15. E. Giacoumidis, J. L. Wei, X. Q. Jin, and J. M. Tang, "Improved transmission performance of adaptively modulated optical OFDM signals over directly modulated DFB laser-based IMDD links using adaptive cyclic prefix", Opt. Exp., vol. 16, no. 13, pp. 9480–9494, 2008.

16. R. P. Giddings et al., "Experimental demonstration of record high 11.25Gb/s real-time end-to-end optical OFDM transceivers for PONs", Opt. Exp., vol. 18, no. 6, pp. 5541-5555, 2010.

17. E. Giacoumidis, M. A. Jarajreh, S. Sygletos, S. T. Le, A. Tsokanos, A. Hamié, E. Pincemin, Y. Jaouën, F. Farjady, A. D. Ellis, and N. J. Doran, "Dual-polarization multi-band OFDM transmission and transceiver limitations for up to 500 Gb/s in uncompensated long-haul links", Opt. Exp., vol. 22, no. 9, pp. 10975–10986, 2014.

18. Y. Lin, E. Giacoumidis, S. O'Duill, and L. P. Barry, "SOA-based wavelength conversion of a coherent optical 64-QAM signal", In Proc. Photonics Ireland (PI), 3-5 September 2018, Cork, Ireland.

19. Y. Lin, E. Giacoumidis, S. O'Duill, A. Arvind, and L. P. Barry, "Reduction of nonlinear distortion in SOA-based wavelength conversion system by post-blind-compensation based on machine learning clustering", In Proc. OFC, 3-7 March 2019, CA, USA (accepted)

20. E. Giacoumidis, S. K. Ibrahim, J. Zhao, J. L. Wei, J. M. Tang, A. D. Ellis, and I. Tomkos, "Effect of ADC on the Performance of Optical Fast-OFDM in MMF/SMF-Based Links", In Proc. PIERS, 12-16 September 2011, Suzhou, China, pp. 402-406, ISBN: 978-1-934142-18-9.

21. E. Giacoumidis, S. K. Ibrahim, J. Zhao, J. M. Tang, A. D. Ellis, and I. Tomkos, "Experimental Demonstration of Cost-Effective Intensity-Modulation and Direct-Detection Optical Fast-OFDM over 40km SMF Transmission", In Proc. OFC/NFOEC, 4-8 March 2012, Los Angeles, CA, USA, p. JW2A.65.

22. E. Giacoumidis, S. K. Ibrahim, J. Zhao, J. M. Tang, A. D. Ellis, and I. Tomkos, "Experimental and Theoretical Investigations of Intensity-Modulation and Direct-Detection Optical Fast-OFDM over MMF-links", IEEE Phot. Techn. Lett., vol. 24, no. 1, pp. 52−54, 2012.

23. E. Giacoumidis, A. Tsokanos, C. Mouchos, G. Zardas, C. Alves, J. L. Wei, J. M. Tang, C. Gosset, Y. Jaouën, and I. Tomkos, "Extensive Comparisons of Optical Fast-OFDM and Conventional Optical OFDM for Local and Access Networks", J. of Opt. Commun. & Netw., vol. 4, no. 10, pp. 724−733, 2012.

24. J. Karaki et al., "Dual-polarization multi-band OFDM versus single-carrier DP-QPSK for 100 Gb/s long-haul WDM transmission over legacy infrastructure," Opt. Exp., vol. 21, no. 14, pp. 16982−16991, 2013.

## 7. Acknowledgments

This work was emanated from EU Horizon 2020 research and innovation programme under the Marie Skłodowska-Curie grant agreement No 713567 and in part by a research grant from Science Foundation Ireland (SFI) and is co-funded under the European Regional Development Fund under Grant Number 13/RC/2077.